\documentclass[twocolumn,showpacs,preprintnumbers,amsmath,amssymb]{revtex4}

\usepackage{graphicx}
\usepackage{dcolumn}
\usepackage{bm}
\newcommand{\Eq}[1]{&\hspace{-0.5em}#1\hspace{-0.5em}&}

\begin{document}


\title{SU(3) dibaryons in the Einstein-Skyrme model}

\author{Hisayuki Sato$^{1}$}
 \email{sugar@dec.rikadai.jp}
\author{Nobuyuki Sawado$^{1}$}
 \email{sawado@ph.noda.tus.ac.jp}

\affiliation{
$^{1}$Department of Physics, Faculty of Science and Technology, 
Tokyo University of Science, Noda, Chiba 278-8510, Japan
}

\date{\today}

\begin{abstract}
SU(3) collective coordinate quantization 
to the regular solution of the $B=2$ 
axially symmetric Einstein-Skyrme system is performed. 
For the symmetry breaking term, a perturbative treatment as well as 
the exact diagonalization method called Yabu-Ando approach are used. 
The effect of the gravity on the mass spectra of the SU(3) dibaryons and the 
symmetry breaking term is studied in detail. In the strong gravity limit, 
the symmetry breaking term significantly reduces and exact SU(3) flavor 
symmetry is recovered. 
\end{abstract}

\maketitle

\section{Introduction}
The Skyrme model~\cite{sky} is considered as an unified theory of hadrons 
by incorporating baryons as topological solitons of pion fields, called skyrmions. %
The topological charge is identified as the baryon number $B$. %
Performing collective quantization for a $B=1$ skyrmion, one can obtain 
the proton and the neutron within 30$\%$ error~\cite{ad83}. %
Correspondingly multi-skyrmion ($B>1$) solutions are expected to represent 
nuclei~\cite{bra,manton}.

The Einstein-Skyrme (ES) model can be thought of as a model of hadrons 
in which baryons interact with black holes.  
The early studies \cite{moss,droz,bizon} have shown that the Schwarzschild black 
hole can support spherically symmetric (hedgehog) Skyrme hair
which is a first counter example to the no-hair conjecture. 
Axially symmetric regular and black hole skyrmion solutions with $B=2$ were 
found in~Ref.\cite{np}. Also some multi-skyrmion solutions $B>2$ have been 
obtained \cite{ioannidou}.

If a skyrmion is regarded as a nucleon or a nuclei, 
it must be quantized to assign quantum numbers like spin, isospin, {\it etc.}. 
Study of gravitational effects on the quantum spectra of skyrmions was 
initiated in Ref.~\cite{np2} by performing collective quantization of a $B=1$ 
gravitating skyrmion. The study about a $B=2$ axially symmetric skyrmion 
immediately followed to it~\cite{myfp}.
It was shown there that the qualitative change in the mass difference, mean charge 
radius and densities under the strong gravitational influence confirms the 
attractive feature of gravity while the reduction of the axial coupling 
and transition moments indicates the gravitational effects as 
a stabilizer of baryons.

In Ref.~\cite{myfp}, we investigated the gravity effects to the SU(2)
solutions of the ES model and found out that the effects are seen especially in 
the heavier dibaryon spectra.  
In this paper we shall study the gravity effects to the SU(3) dibaryons  
because they have much larger mass compared with that of SU(2) 
dibaryons in terms of the large flavor symmetry breaking.
The effect will be more apparent on such dihyperons.

The extension of the Skyrme model into SU(3) have extensively been studied.
Roughly speaking, solutions of the SU(3) are classified into two categories:
one is embedded, another is non-embedded ones. 
For the non-embedded solutions, in Ref.\cite{kleihaus} the classical SO(3) chiral field 
is simply extended into three flavor space. The authors constructed the solutions 
in the ES system and found particle-like and black hole solutions. 
More systematic analysis in this direction has been done in Refs.\cite{ioannidou,brihaye}. 
By using the harmonic map ansatz, the authors have investigated various large flavor and 
multi winding number solitons.

On the other hand, for the embedded solutions, the successful approaches are based on perturbations. 
In the bound state approach \cite{callan} static baryons are considered as bound states of 
the kaon and the skyrmions. The chiral field is expanded in the kaon fluctuations around the classical solutions.  
This approximation works well for the large symmetry breaking. 
Collective coordinate approach is essentially a natural extension of the SU(2) collective quantization scheme, 
including symmetry breaking terms. Good quantum numbers for spin and isospin are 
obtained by quantizing the their rotational zero modes \cite{guadagnini, mazur}. 
Yabu-Ando (YA) treatment \cite{yabu} is a sort of the collective quantization but the collective Hamiltonian 
is exactly diagonalized by using Euler-angle parameterization of the SU(3) rotations. 
As a result, baryon appears to be its lowest irreducible representation (irrep) 
but contain significant admixture of higher representations. 
The application to the $B=2$ axially symmetric skyrmions \cite{kope1992} and to 
the multi-skyrmions ($3\le B\le 8$) \cite{schat} in flat space have been studied. They observed 
large mixing of higher irreps especially in excited or higher winding 
number states. 

In this Letter, we construct SU(3) axially symmetric skyrmions in the ES system in terms of the 
collective quantization and YA approach. 
We estimate mass spectra of the dibaryons down to strangeness $S=-6$.
Since our concern is about contributions of higher irreps to the lowest states 
and we shall show change of the mixing probabilities about varying coupling strength of the gravity.

\begin{figure*}
\begin{tabular}{c@{\hspace{5mm}}c}
\includegraphics[width=9cm,keepaspectratio,clip]{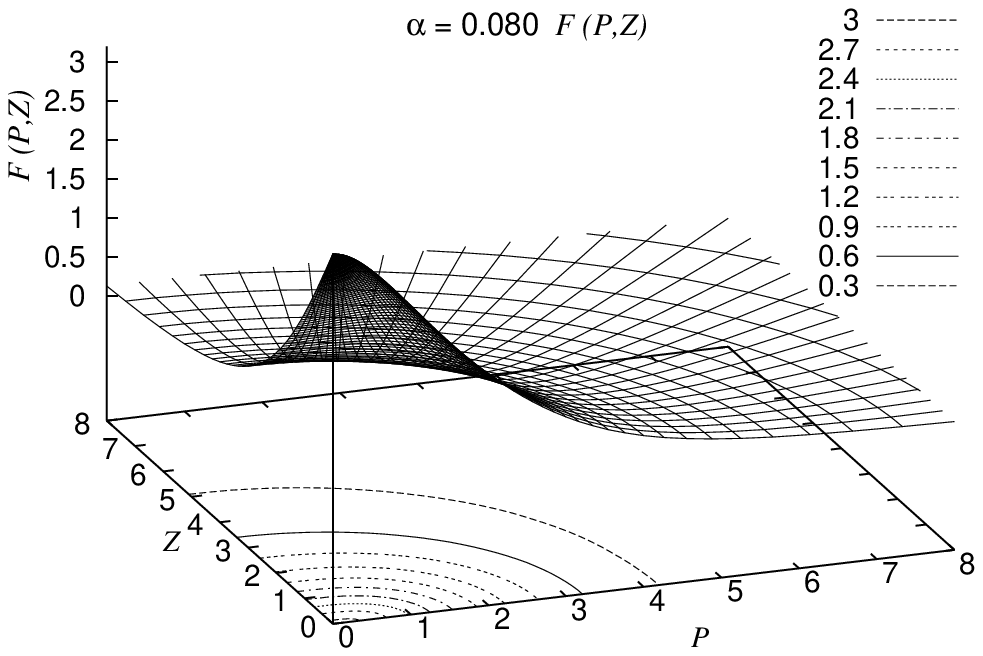}
\includegraphics[width=9cm,keepaspectratio,clip]{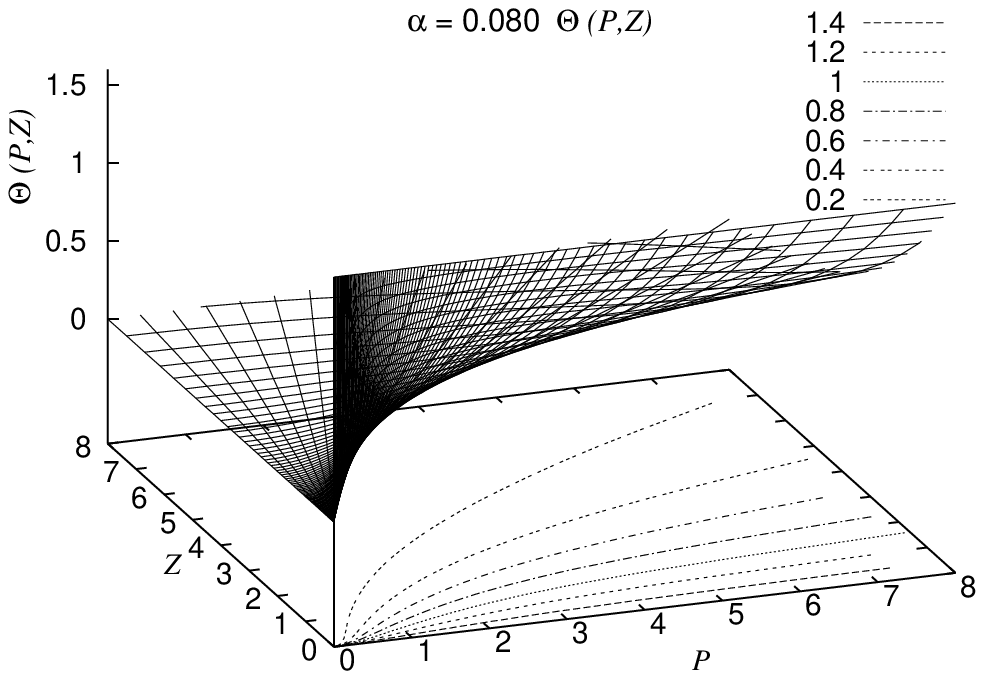}
\end{tabular}
\caption{\label{fig1} The profile functions $F,\Theta$ for $\alpha=0.080$ in the cylindrical coordinate with 
dimensionless variables $P:=ef_\pi \rho,Z=ef_\pi z$.}
\end{figure*}

\begin{figure}
\begin{tabular}{c@{\hspace{5mm}}c}
\includegraphics[width=8.5cm,keepaspectratio,clip]{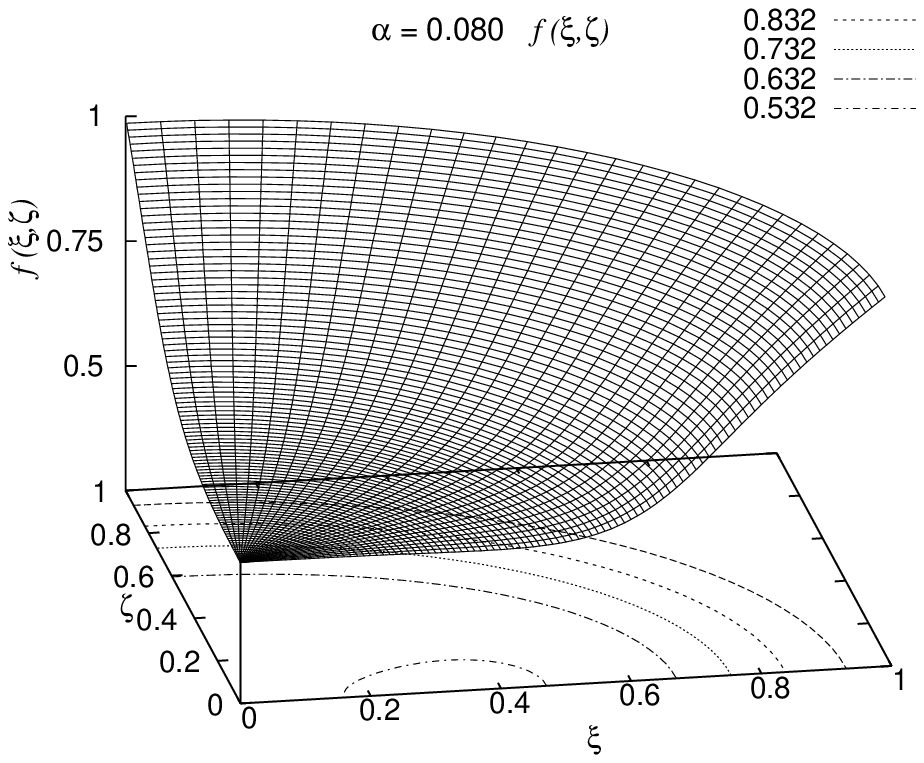}\\
\includegraphics[width=8.5cm,keepaspectratio,clip]{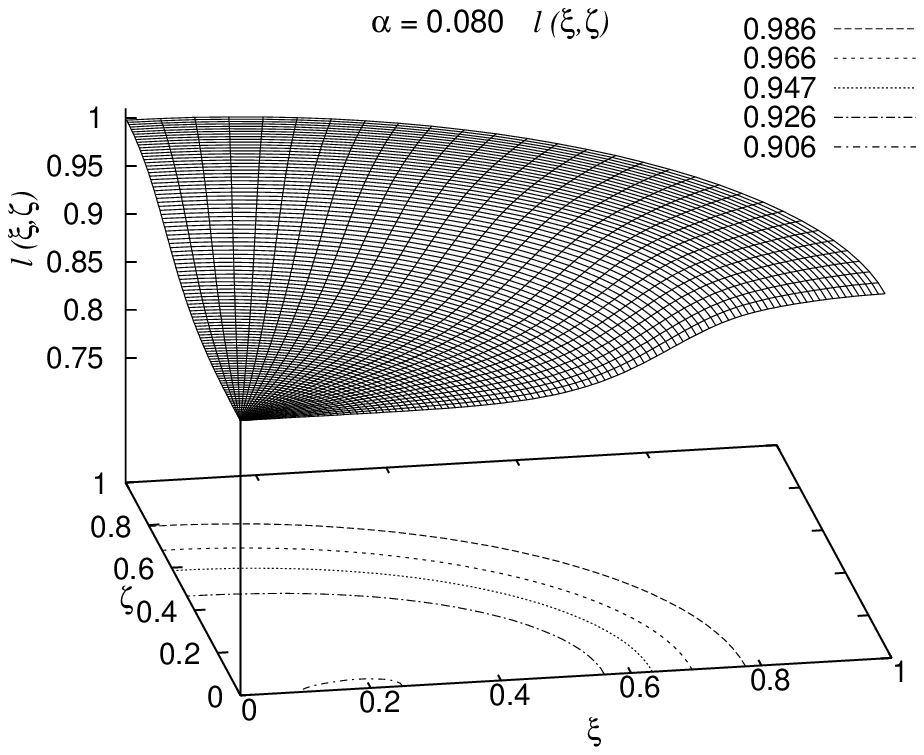}\\
\includegraphics[width=8.5cm,keepaspectratio,clip]{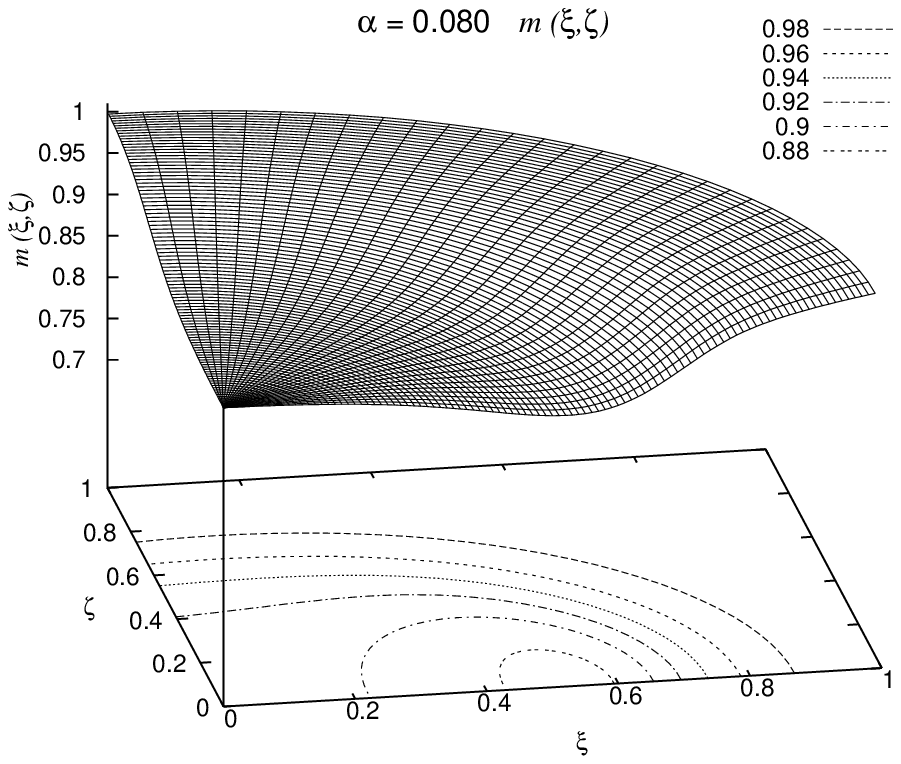}
\end{tabular}
\caption{\label{fig2} The metric functions $f,l,m$ in the cylindrical coordinate with
dimensionless, rescaled variables $\xi=P/(1+P), \zeta=Z/(1+Z), P=ef_\pi\rho,Z=ef_\pi z$.}
\end{figure}

\begin{center}
\begin{figure*}
\begin{tabular}{c@{\hspace{5mm}}c}
\begin{minipage}{80mm}
\begin{center}
\end{center}
\begin{center}
\includegraphics[width=7.5cm,keepaspectratio,clip]{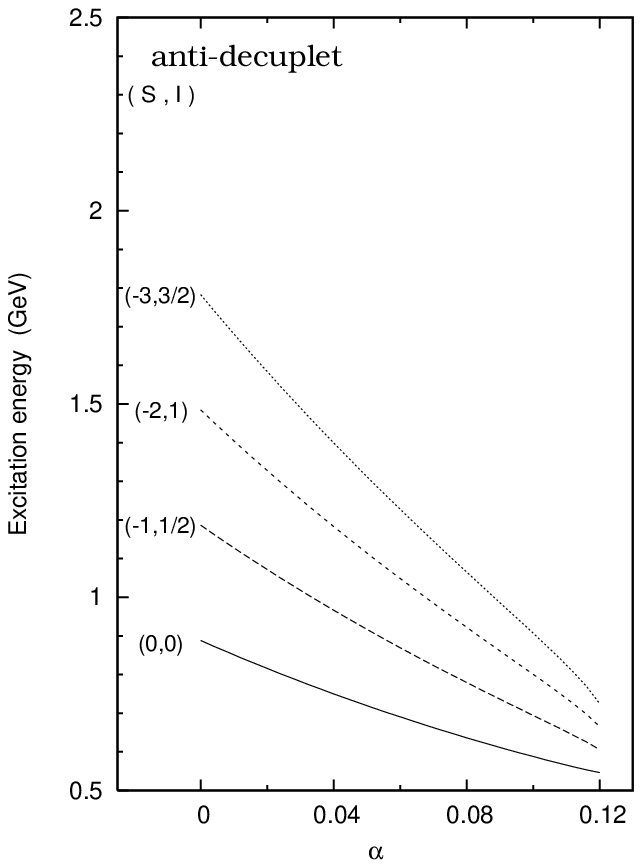}
\end{center}
\end{minipage}
&
\begin{minipage}{80mm}
\begin{center}
\includegraphics[width=7.5cm,keepaspectratio,clip]{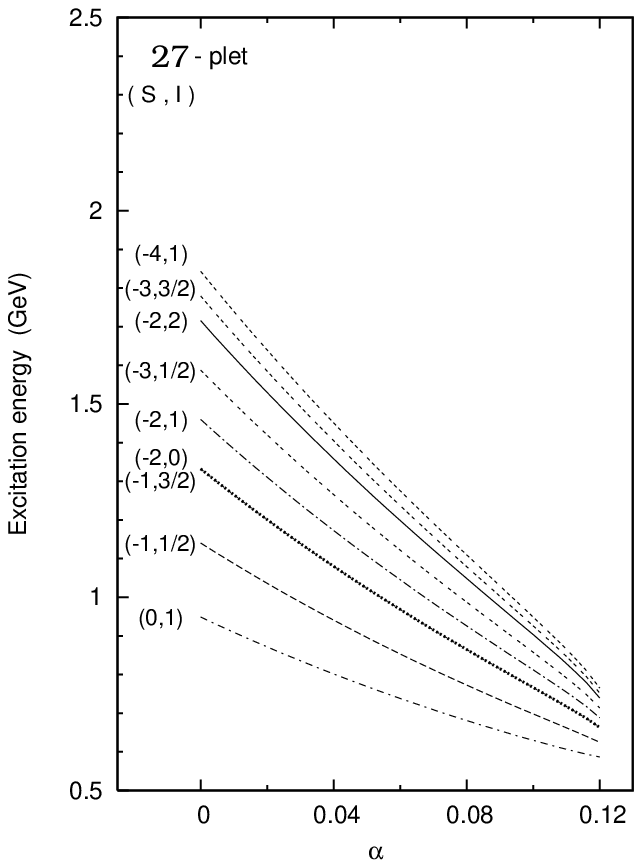}
\end{center}
\end{minipage}\\

\begin{minipage}{80mm}
\begin{center}
\includegraphics[width=7.5cm,keepaspectratio,clip]{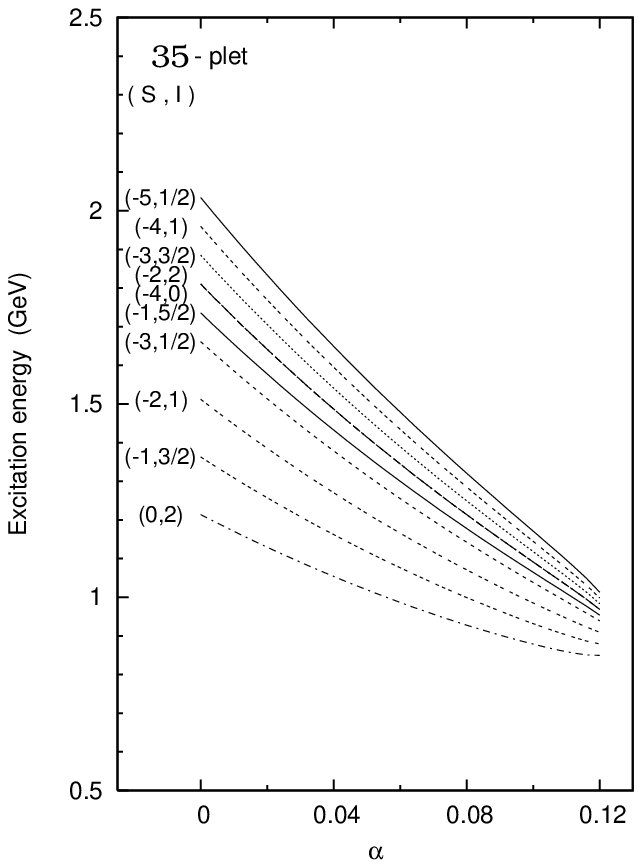}
\end{center}
\end{minipage}
&
\begin{minipage}{80mm}
\begin{center}
\includegraphics[width=7.5cm,keepaspectratio,clip]{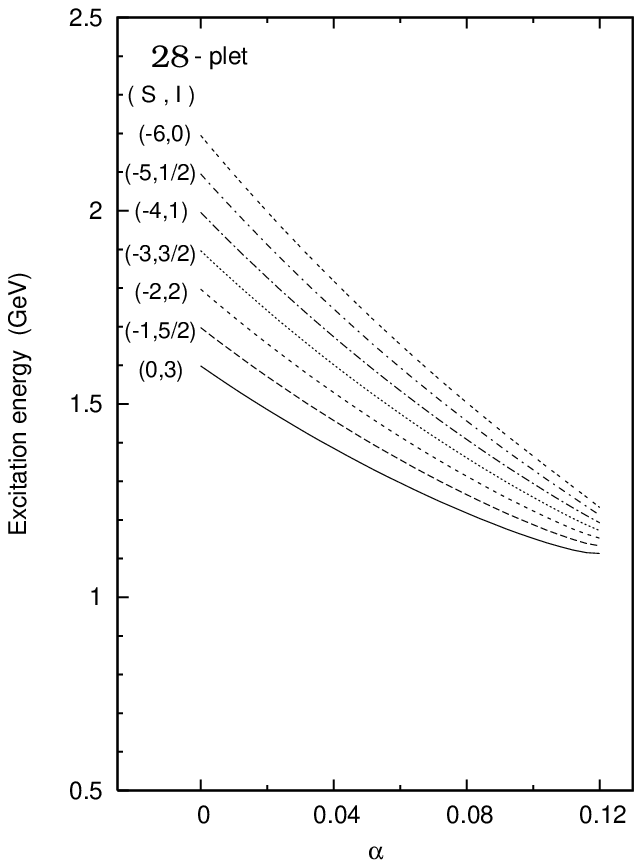}
\end{center}
\end{minipage}
\end{tabular}
\caption{\label{fig3} The coupling constant dependence of the 
mass difference from the classical energy 
for the multiplets $\{ \overline{10} \}$,$\{27\}$, $\{35\}$,$\{28\}$ 
($(p,q)=(0,3),(2,2),(4,1),(6,0)$,respectively) 
are shown in the unit of GeV. The quantized energy is computed
in terms of the naive perturbation technique. }
\end{figure*}
\end{center}

\begin{center}
\begin{figure*}
\begin{tabular}{c@{\hspace{5mm}}c}
\begin{minipage}{80mm}
\begin{center}
\includegraphics[width=7.5cm,keepaspectratio,clip]{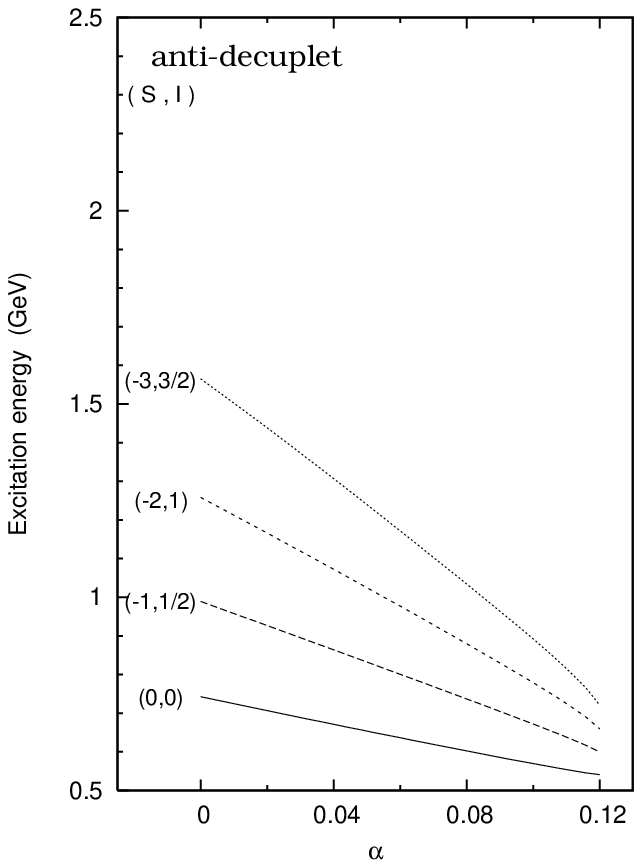}
\end{center}
\end{minipage}
&
\begin{minipage}{80mm}
\begin{center}
\includegraphics[width=7.5cm,keepaspectratio,clip]{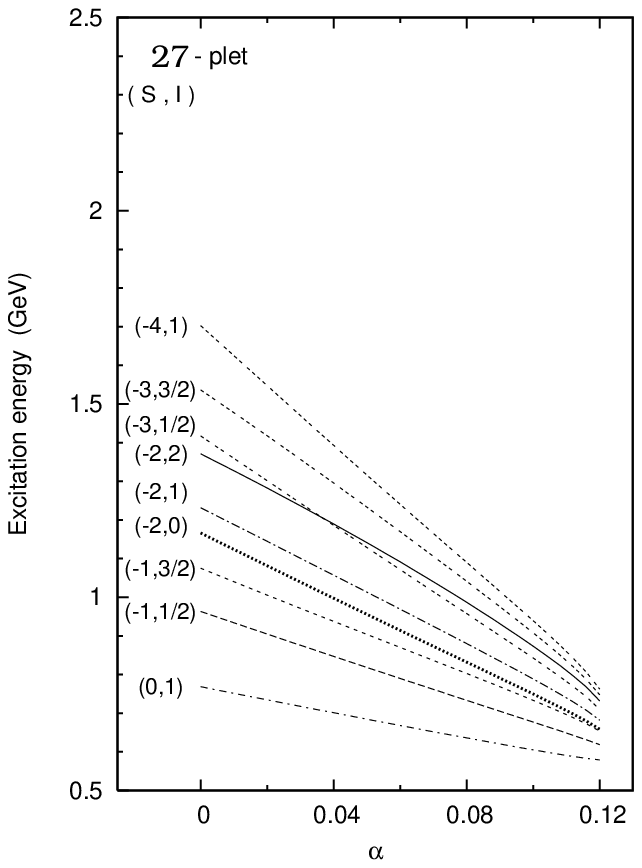}
\end{center}
\end{minipage}\\

\begin{minipage}{80mm}
\begin{center}
\includegraphics[width=7.5cm,keepaspectratio,clip]{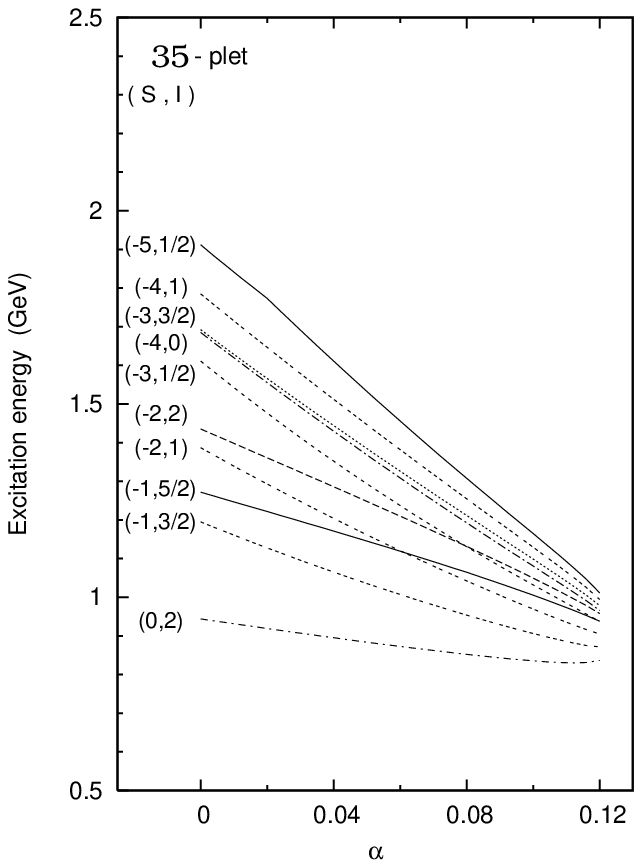}
\end{center}
\end{minipage}
&
\begin{minipage}{80mm}
\begin{center}
\includegraphics[width=7.5cm,keepaspectratio,clip]{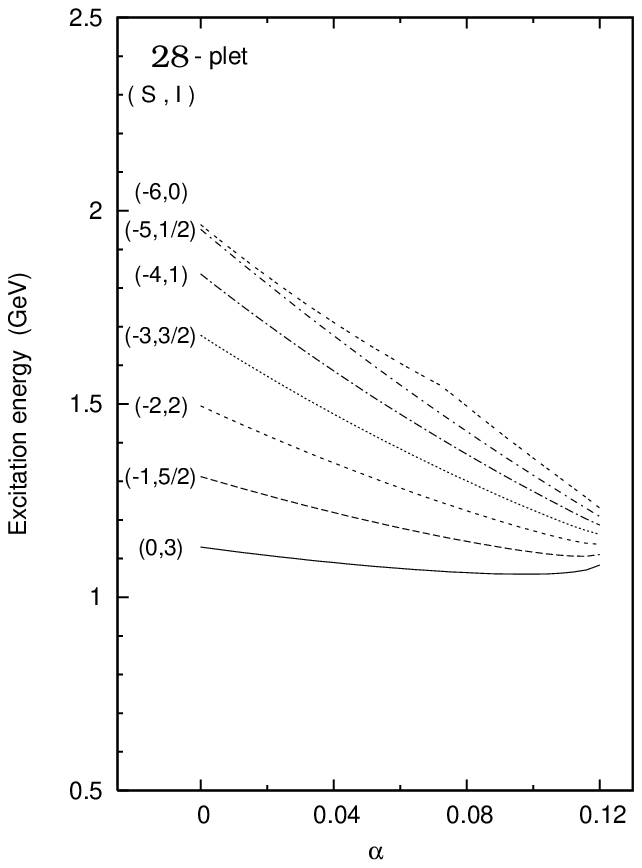}
\end{center}
\end{minipage}
\end{tabular}
\caption{\label{fig4}  The coupling constant dependence of the 
mass difference from the classical energy 
for the multiplets $\{ \overline{10} \}$,$\{27\}$, $\{35\}$,$\{28\}$ 
($(p,q)=(0,3),(2,2),(4,1),(6,0)$,respectively) 
are shown in the unit of GeV. The quantized energy is computed
in terms of YA. }
\end{figure*}
\end{center}

\begin{figure}
\begin{center}
\includegraphics[width=9cm,keepaspectratio,clip]{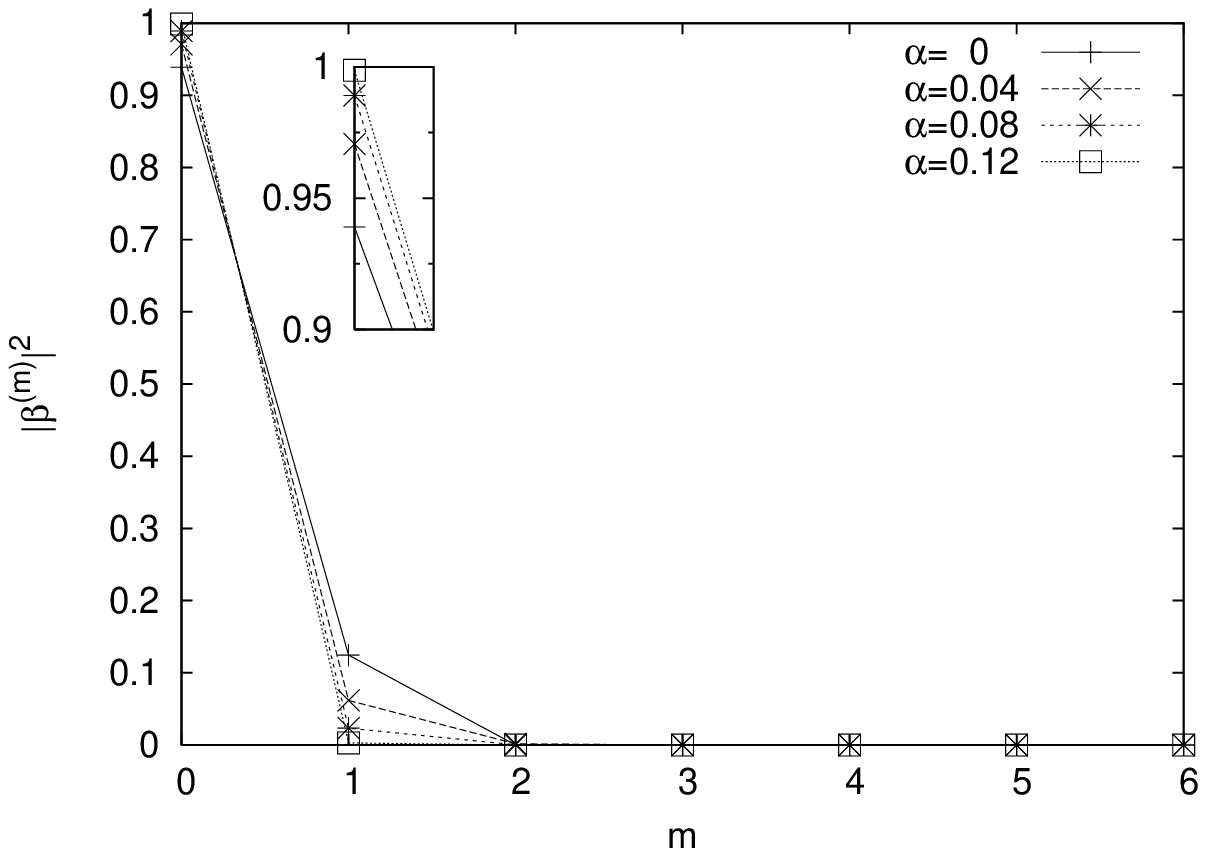}
\end{center}
\caption{\label{fig5}Mixing probability of the higher representations to 
the lowest energy state of $NN$ channel in $\{\overline{10}\}$. }
\end{figure}

\section{CLASSICAL GRAVITATING $B=2$ SKYRMIONS}
In this section we shall introduce basic formalism and classical solutions 
of the axially symmetric gravitating skyrmions. 
The SU(3) extended Skyrme Lagrangian coupled with gravity 
can be written as
\begin{eqnarray}
L =L_G+L_S+L_{SB}+L_{WZ} \label{Lag}
\end{eqnarray}
where $L_G$ is the standard Einstein-Hilbert Lagrangian
\begin{eqnarray}
&&L_G =\int d^3r  \sqrt{-g} \frac{1}{16\pi G}R\label{GLe}
\end{eqnarray}
and the remaining parts are about the SU(3) extension of the Skyrme Lagrangian, which are 
defined in terms of the chiral field $U(x)\in$ SU(3) and using the notation 
$l_\mu=U^\dagger\partial_\mu U$ as follows:
\begin{eqnarray}
&&L_S =\int d^3r \sqrt{-g} \Bigl[ \frac{1}{16}F_{\pi}^2g^{\mu \nu}
{\rm Tr}(l_\mu l_\nu)\nonumber \\
&&~~~~+\frac{1}{32e^2}g^{\mu \rho}g^{\nu \sigma}
{\rm Tr}([l_\mu,l_\nu][l_\rho,l_\sigma) \Bigl] ,
\label{lag_skyrme} \\
&&L_{SB}=\frac{1}{16}F_{\pi}^2m_{\pi}^2 \int d^3r \sqrt{-g}
{\rm Tr}(U+U^{\dagger}-2) \nonumber \\
&&~~~~+\frac{1}{24}(F_{\kappa}^2m_{\kappa}^2-F_{\pi}^2m_{\pi}^2)\int d^3r\sqrt{-g}
{\rm Tr}(1-\sqrt{3}\lambda_8) \nonumber \\
&&~~~~\times (U+U^{\dagger}-2) , 
\label{lag__sb}\\
&&L_{WZ}=-\frac{iN_c}{240\pi^2}\int_Q d\Sigma^{\mu \nu \lambda \rho \sigma}
{\rm Tr}[ l_\mu l_\nu l_\lambda l_\rho l_\sigma],
\label{lag_wz}
\end{eqnarray}
where $F_{\pi}$ and $e$ are basic model parameters which indicate the pion decay constant, 
a dimensionless parameter, respectively. 
The $L_{SB}$ is comprised of all chiral and flavor symmetry breaking terms. 
The $L_{WZ}$ is usual Wess-Zumino term which concerns with the topological 
charge of the soliton. $F_{\kappa}, m_\kappa$ are the kaon decay constant 
and the mass. $N_c$ means a number of color.

In this Letter, we employ the solution of the chiral field with axially symmetric, winding number two 
which is considered as a possible candidate for the $B=2$ minimal energy configuration~\cite{bra}
\begin{eqnarray}
&&U_0(\bm{r})={\rm exp}[iF(r,\theta)\bm{\tau}\cdot \bm{n_R}]. \label{ansatz_pir}
\end{eqnarray}
where $\bm{n_R}$ is defined as
\begin{eqnarray}
\bm{n_R}=(\sin{\Theta}(r,\theta)\cos{n\varphi},\sin{\Theta}(r,\theta)\sin{n\varphi},\cos{\Theta}(r,\theta))
\label{ansatz_pit}
\end{eqnarray}
and $n \in \mathbb{Z}$ is the winding number. We explore the solution with $n=2$. 
SU(3) chiral field is constructed by trivial embedding:
\begin{eqnarray}
&&U(\bm{r})=\left(
	\begin{array}{cc}
	U_0(\bm{r})& 0 \\
	0 & 1 \\
	\end{array}\right)\,.
	\label{ansatz_su3} 
\end{eqnarray}

Correspondingly, the following axially symmetric ansatz is imposed on the 
metric~\cite{kunz}
\begin{eqnarray}
	ds^2=-fdt^2+\frac{m}{f}(dr^2+r^2d \theta^2)+\frac{l}{f}r^2 \sin^2 \theta d \varphi^2 
\end{eqnarray}
where the metric functions $f$ , $m$ and $l$ are the function of coordinates $r$ and $\theta$. 
This metric is symmetric with respect to the $z$-axis ($\theta=0$). 
Substituting these ansatz to the Lagrangian (\ref{Lag}), 
one obtains the following static (classical) energy for the chiral fields
\begin{eqnarray}
&&M_{class}=2\pi {\textstyle \frac{F_{\pi}}{e}} \int dxd\theta 
\Bigl[ {\textstyle \frac{\sqrt{l}\sin{\theta}}{8}} \nonumber \\
&&~~~~\times \bigl \{ x^2 \left( (\partial_{x}F)^2+(\partial_{x}\Theta)^2\sin^2{F} \right) \nonumber \\
&&~~~~+(\partial_{\theta}F)^2+(\partial_{\theta}\Theta)^2\sin^2{F}
	+{\textstyle \frac{n^2m}{l \sin{\theta}}}\sin^2{F}\sin^2{\Theta} \bigr\} \nonumber \\
&&~~~~+{\textstyle \frac{\sqrt{l}\sin{\theta}}{2}} \Bigl[ {\textstyle \frac{f}{m}}
	(\partial_{x}F\partial_{\theta}\Theta-\partial_{\theta}F\partial_{x}\Theta)^2\sin^2{F} \nonumber \\
&&~~~~+{\textstyle \frac{n^2f}{l\sin^2{\theta}}}\sin^2{F}\sin^2{\Theta} \bigl\{ 
\bigl((\partial_{x}F)^2+{\textstyle \frac{1}{x^2}}(\partial_{\theta}F)^2\bigr)\nonumber \\
&&~~~~+\bigl((\partial_{x}\Theta)^2+{\textstyle \frac{1}{x^2}}(\partial_{\theta}\Theta)^2\bigr)\sin^2{F}
 \bigr\} \Bigr]\, \nonumber \\
&&~~~~+{\textstyle \frac{1}{4}\frac{m\sqrt{l}}{f}}x^2 \sin{\theta}\beta_{\pi}^2(1-\cos{F}) \Bigr],
 \label{edene}
\end{eqnarray}
where dimensionless variable $x=e F_{\pi}r$ 
 and $\beta_{\pi}=\frac{m_{\pi}}{eF_{\pi}}$ are introduced.

For the profile functions, the boundary conditions at the $x=0, \infty$ are imposed
\begin{eqnarray}
&&F(0,\theta )= \pi , \ F(\infty,\theta )=0 , \\
&&\partial_{x}\Theta (0,\theta)= \partial_{x}\Theta (\infty,\theta)=0 . 
\end{eqnarray}
At $\theta = 0$ and $\pi/2$, 
\begin{eqnarray}
&&\partial_{\theta}F(x,0)=\partial_{\theta}F(x,\frac{\pi}{2})=0 , \\
&&\Theta (x,0)=0  \ , \  \Theta (x,\frac{\pi}{2})=\frac{\pi}{2} .
\end{eqnarray}
For the solutions to be regular at the origin $x=0$ and to be asymptotically flat at infinity, 
the following boundary conditions must be imposed 
\begin{eqnarray}
&&\partial_{x}f(0,\theta)=\partial_{x}m(0,\theta)=\partial_{x}l(0,\theta)=0 , \label{mbr0e} \\
&&f(\infty ,\theta)=m(\infty ,\theta)=l(\infty ,\theta)=1 . \label{mbr1e}
\end{eqnarray}
For the configuration to be axially symmetric, the following boundary conditions 
must be imposed at $\theta = 0$ and $\pi/2$ 
\begin{eqnarray}
&&\partial_{\theta}f(x,0)=\partial_{\theta}m(x,0)=\partial_{\theta}l(x,0)=0 , \label{mba0e} \\
&&\partial_{\theta}f(x,\frac{\pi}{2})=\partial_{\theta}m(x,\frac{\pi}{2})=\partial_{\theta}l(x,\frac{\pi}{2})=0 . \label{mba1e} 
\end{eqnarray}

The covariant topological current is defined by
\begin{eqnarray}
B^{\mu}= \frac{\epsilon^{\mu \nu \rho \sigma}}{24 \pi^2} \frac{1}{\sqrt{-g}}
	{\rm tr}(U^{-1}\nabla_{\nu}UU^{-1}\nabla_{\rho}UU^{-1}\nabla_{\sigma}U). \label{bcde}
\end{eqnarray}
Substituting the ansatz (\ref{ansatz_pir}),(\ref{ansatz_pit}) into (\ref{bcde})
the zeroth component is estimated as
\begin{eqnarray}
B^0 = -\frac{1}{\pi ^2\sqrt{-g}} \sin^2{F} \sin{\Theta}
 (\partial_{x} F \partial_{\theta}\Theta-\partial_{\theta} F \partial_{x}\Theta  ). 
\end{eqnarray}
The baryon number of the soliton $B$ is derived from its spatial integral, thus
\begin{eqnarray}
	B\Eq{=}\int d^3r \sqrt{-g} B^{0} \nonumber \\
	\Eq{=} \frac{1}{2\pi}(2F-\sin{2F})\cos{\Theta} \biggr|_{F_0,\Theta_0}^{F_1,\Theta_1}. 
\end{eqnarray}
The inner and outer boundary conditions $(F_0,\Theta_0)=(\pi,0)$ and $(F_1,\Theta_1)=(0,\pi)$ yield $B=2$. 

By taking a variation of the static energy (\ref{edene}) with respect to $F$ and $\Theta$, 
one obtains the equations of motion for the profile functions. 
The field equations for the metric functions $f$, $m$ and $l$ are derived 
from the Einstein equations. The explicit form of the equations is essentially 
same (except for contribution of the mass term) as reported in Ref.\cite{myfp}.

The effective coupling constant of the Einstein-Skyrme system is given by 
\begin{eqnarray}
	\alpha = 4 \pi G F_{\pi}^2 
\end{eqnarray}
which is the only free parameter. 

To solve the equations of motion numerically, the relaxation method with the typical 
grid size $100 \times 30$ are performed. We observe that the solution 
survives at $0 \leq \alpha \leq 0.120$. Including the mass term, 
the range becomes a little narrow. We show examples of our numerical results for the 
profile functions $F,\Theta$ in Fig.\ref{fig1} and also for the metric 
functions $f,l,m$ in Fig.\ref{fig2}.

\section{THE SU(3) COLLECTIVE QUANTIZATION}
We study the SU(3) extension of the axially symmetric $B=2$ skyrmions 
by Yabu-Ando approach together with naive collective coordinate quantization. 
SU(3) chiral field is constructed by trivial embedding 
\begin{eqnarray}
	\tilde{U}(\bm{r},t)=A(t)\left(
	\begin{array}{cc}
	U_0(R(t) \bm{r})& 0 \\
	0 & 1 \\
	\end{array}\right)
	A^{\dagger}(t) 
	\label{ansatz_csu3}
\end{eqnarray}
where $U_0$ is introduced in Eq.(\ref{ansatz_pir}). $A(t)$ is time dependent SU(3) rotational matrix and 
$R(t)$ describes a spatial rotation of the soliton. 
We introduce the angular velocities $\Omega_a,\omega_l$ which are defined by
\begin{eqnarray}
&&A^{\dagger}\dot{A}=\frac{i}{2}\sum_{a=1}^8\lambda_a \Omega_a\,,  \\
&&(\dot{R}R^{\dagger})_{ik}=\sum_{l=1}^3\varepsilon_{ikl}\omega_l\,. \label{}
\end{eqnarray}
Substituting the chiral field (\ref{ansatz_csu3}) into the Lagrangians (\ref{lag_skyrme})-(\ref{lag_wz}) and 
after a lengthy calculation, one finally obtain the effective Lagrangian of the form:
\begin{eqnarray}
&&L=-M_{class}+\frac{1}{2}I_{N} \sum_{p=1}^2\Omega_p^2
	+\frac{1}{2}I_{J}\sum_{p=1}^2\omega_p^2 \nonumber \\
&&~~~~+\frac{1}{2}I_{3}(\Omega_3^2+n\omega_3)^2
	+\frac{1}{2}I_{S}\sum_{k=4}^7\Omega_k^2 \nonumber \\
&&~~~~-\frac{N_c}{2\sqrt{3}}\Omega_8+\frac{1}{2}\gamma(1-D_{88}(A)), \label{Lag2}
\end{eqnarray}
where $I_N,I_J,I_3,I_S$ are called the moments of inertia and their explicit 
forms are  
\begin{eqnarray}
I_{N} \Eq{=}{\textstyle \frac{\pi}{F_{\pi}e^3}} \int dx d\theta 
			\Bigl[ {\textstyle \frac{m\sqrt{l}}{4f^2}}x^2 \sin{\theta}
						\sin^2{F}(1+\cos^2{\Theta}) \nonumber \\ 
&&+{\textstyle \frac{\sqrt{l}}{f}}\sin{\theta}\sin^2{F} 
   \bigl\{(1+\cos^2{\Theta})(x^2(\partial_{x}F)^2+(\partial_{\theta}F)^2) \nonumber \\
&&+\sin^2{F}\cos^2{\Theta}(x^2(\partial_{x}\Theta)^2+(\partial_{\theta}\Theta)^2) \nonumber \\
&&+{\textstyle \frac{n^2m}{l \sin^2{\theta}}}\sin^2{F}\sin^2{\Theta} \bigr \} \Bigr], \label{thene} \\
I_{3} \Eq{=}{\textstyle \frac{\pi}{F_{\pi}e^3}} \int dx d\theta 
			\Bigl[ {\textstyle \frac{m\sqrt{l}}{2f^2}}x^2\sin{\theta}
			\sin^2{F}\sin^2{\Theta} \nonumber \\
&&+{\textstyle \frac{2\sqrt{l}}{f}}\sin{\theta}\sin^2{F}\sin^2{\Theta}
	 \bigl\{x^2(\partial_{x}F)^2+(\partial_{\theta}F)^2 \nonumber \\
&&+\sin^2{F}(x^2(\partial_{x}\Theta)^2+(\partial_{\theta}\Theta)^2)
	\bigr \} \Bigr], \label{the3e} \\
I_{J} \Eq{=}{\textstyle \frac{\pi}{F_{\pi}e^3}} \int dx d\theta
			\Bigl[ {\textstyle \frac{m\sqrt{l}}{4f^2}}x^2\sin{\theta}
			((\partial_{x}F)^2+(\partial_{\theta}\Theta)^2\sin^2{F} \nonumber \\
&& \hspace{15mm}  +n^2\cot^2{\theta}\sin^2{F}\sin^2{\Theta}) \nonumber \\
&&+{\textstyle \frac{\sqrt{l}}{f}}x^2\sin{\theta}\sin^2{F}
	\bigl\{(\partial_{x}F\partial_{\theta}\Theta-\partial_{\theta}F\partial_{x}\Theta)^2 \nonumber \\
&& \hspace{10mm}+n^2((\partial_{x}F)^2
	+(\partial_{x}\Theta)^2\sin^2{F})\cot^2{\theta}\sin^2{\Theta} \bigr \} \nonumber \\
&&+{\textstyle \frac{n^2 \sqrt{l}}{f \sin{\theta}}}(\cos^2{\theta}+{\textstyle \frac{m}{l}}) \nonumber \\
&& \hspace{5mm}\times ((\partial_{\theta}F)^2+(\partial_{\theta}\Theta)^2\sin^2{F})
	 \sin^2{F} \sin^2{\Theta} \Bigr], \label{theje}\\
I_{S} \Eq{=}{\textstyle \frac{\pi}{F_{\pi}e^3}}\int dxd\theta 
	(1-\cos{F})\Bigl[{\textstyle \frac{m\sqrt{l}}{4f^2}}x^2\sin{\theta} \nonumber \\
&& +{\textstyle \frac{\sqrt{l}}{4f}}x^2\sin{\theta}
	\{ (\partial_{x}F)^2+\sin^2{F}(\partial_{x}\Theta)^2 \} \nonumber \\
&& +{\textstyle \frac{\sqrt{l}}{4f}}\sin{\theta}
	\{ (\partial_{\theta}F)^2+\sin^2{F}(\partial_{\theta}\Theta)^2 \nonumber \\
&& \hspace{10mm}+{\textstyle \frac{mn^2}{l\sin^2{\theta}}} \sin^2{F}\sin^2{\Theta} \} \Bigr].\label{}
\end{eqnarray}
$\frac{1}{2}\gamma(1-D_{88})$ exhibits strength of the symmetry breaking and 
the explicit form of $\gamma$ is 
\begin{eqnarray}
\gamma ={\textstyle \frac{2\pi F_{\pi}}{3e}}(\beta_{\kappa}^2-\beta_{\pi}^2)\int
dxd\theta x^2\sin{\theta}(\cos{F}-1), \label{}
\end{eqnarray}
where $\beta_{\kappa}=\frac{m_{\kappa}F_{\kappa}}{eF_{\pi}^2}$. 
$D_{88}$ is a component of Wigner function which is defined as 
\begin{eqnarray}
D_{ab}(A)=\frac{1}{2}{\rm Tr}(\lambda_aA^{\dagger}\lambda_bA). \label{}
\end{eqnarray}
From (\ref{Lag2}) the Hamiltonian reads
\begin{eqnarray}
H\Eq{=}M_{class}+\frac{J(J+1)}{2I_{J}}+\frac{1}{2}\Bigl(\frac{1}{I_{N}}-\frac{1}{I_{S}}\Bigr)N(N+1)\nonumber \\
&&+\frac{1}{2}\Bigl(\frac{1}{I_{3}}-\frac{1}{I_{N}}-\frac{n^2}{I_{J}}\Bigr)L^2-\frac{3}{8I_{S}}B^2
+\frac{1}{2I_S}\varepsilon_{SB}\label{ham}
\end{eqnarray}
where eigenvalues of diagonal operators are already inserted. 
Here, the eigenvalue of $J$ is spin, $I$ is isospin, 
$N$ is right isospin derived from $N=\frac{1}{2}p_0$ 
where $(p_0,q_0)$ is the minimal irrep 
and $L$ is the third component of the body fixed spin operator 
which determine parity $P$ of the state by the relation of $P=(-1)^L$. 
The eigenvalue of the $\varepsilon_{SB}$ is derived from following eigenequation
\begin{eqnarray}
&&\{C_2[SU(3)]+I_S \gamma (1-D_{88})\}\Psi =\varepsilon_{SB}\Psi
\label{eqsb}
\end{eqnarray}
where $C_2[SU(3)]$ is Casimir operator of SU(3). 

We shall investigate (\ref{eqsb}) in two folds.  
One is to treat the symmetry breaking term perturbatively, 
another is to diagonalize the whole 
via a basis of the SU(3) Wigner functions.

\subsection{Perturbative method}
If symmetry breaking effect $\gamma$ is certainly small, 
perturbative treatment seems to be good approximation. 
We introduce a wave function of the form~\cite{weigel}
\begin{eqnarray}
&&\Psi:=\Phi^{(m)}_{II_3Y,NN_3Y_R,JJ_3}(A){D^{J}}^{*}_{J_3,-nN_3}(R^{-1}) \\
&&\Phi^{(m)}_{II_3Y,NN_3Y_R,JJ_3}(A) 
=\sqrt{d^{(m)}}(-1)^{\frac{Y_R}{2}+N_3} \nonumber \\
&&\times{D^{(m)}}^{*}_{II_3Y,NN_3Y_R}(A^{-1})
\label{base}
\end{eqnarray}
where the dimension of the $(p,q)$ irrep is expressed by 
$d^{(m)}=(p+1)(q+1)(p+q+2)/2$, the $m$ is representation of SU(3) group, 
and subscript of the $Y$ and $Y_R$ is hypercharge and right hypercharge respectively. 

With the operation of the collective quantization, 
the angular velocity $\Omega_8$ appears linear in Eq.(\ref{Lag2}).
Therefore we obtain a constraint 
\begin{eqnarray}
	Y_R=\frac{1}{3}N_cB, \label{}
\end{eqnarray}
which means that the symmetry $U_R(1)$ is redundant. 
Thus we obtain $Y_R=2$.

In terms of Eq.(\ref{base}), the expectation value of the Casimir 
invariants $C_2(SU(3))$ in Eq.(\ref{eqsb}) is easily obtained
\begin{eqnarray}
\langle C_2(SU(3)) \rangle=\frac{1}{3}(p^2+q^2+pq+3(p+q))\,.
\end{eqnarray}
For the symmetry breaking term $I_S\gamma(1-D_{88})$, 
the estimation of the expectation value can be done by performing the 
integral of the three Wigner rotation matrices~\cite{blotz, toyota} 
which is evaluated by the SU(3) Clebsch-Gordan coefficient, 
or the isoscalar factor
\begin{eqnarray}
&&\int dA D_{\nu_3 \nu_3'}^{(m_3)*}(A)D_{\nu_1 \nu_1'}^{(m_1)}(A)D_{\nu_2 \nu_2'}^{(m_2)}(A)\nonumber \\
&&=\frac{1}{d^{(m)}}\sum_{\mu}
\left(\begin{array}{ccc}
m_1 & m_2 & m_{3\mu}\\
\nu_1 & \nu_2 & \nu_3 \\
\end{array} \right)
\left(\begin{array}{ccc}
m_1 & m_2 & m_{3\mu}\\
\nu_1'& \nu_2'& \nu_3' \\
\end{array} \right). 
\label{}
\end{eqnarray}
Computations of the Clebsch-Gordan coefficients can be performed by 
the numerical algorithm of Ref.~\cite{kaeding}.

\subsection{Diagonalization of the collective Hamiltonian}
If symmetry breaking effect is crucial, the naive perturbation will substantially 
fail. Yabu-Ando approach can improves the situation.  
In YA, state of baryon appears to be its lowest irrep but 
contain large admixture of higher irreps. 
We shall see that such mixing reduces in large gravity limit. 

The wave function of the Hamiltonian is expanded in terms of a wave function of the lowest 
representation (\ref{base})
\begin{eqnarray}
&&\Psi:=\sum_m\beta^{(m)}\Phi^{(m)}_{II_3Y,NN_3Y_R,JJ_3}(A){D^{J}}^{*}_{J_3,-nN_3}(R^{-1}) \nonumber \\
\end{eqnarray}
In terms of the basis, the eigenvalue problem in Eq.(\ref{eqsb}) can be reduces to a matrix 
diagonalization problem.

\section{NUMERICAL RESULTS}
For the actual calculations, we fix $F_{\pi}=108 \ {\rm MeV}$, $e=4.84$, $\beta_\pi=0.263$.  
The kaon decay constant, experimentally, is $F_{\kappa}\approx \sqrt{2}F_{\pi}$, 
but for the simplicity, we employ $F_{\kappa}=F_{\pi}$.
For the kaon mass, we employ the experimental value, {\it i.e.}, $\beta_{\kappa}=0.952$. 

We estimate mass spectra belonging to SU(3) multiplets 
$\{ \overline{10} \}$,$\{27\}$,$\{35\}$,$\{28\}$ 
(or in the $(p,q)$ representation, $(0,3),(2,2),(4,1),(6,0)$, 
respectively). The Finkelstein-Rubinstein constraints \cite{FR} 
tells us that for $\{ \overline{10} \}$, $\{35\}$, $J=1$ 
is chosen for the ground state, otherwise one can set $J=0$~\cite{kope2}. 
The eigenvalue of $L$ concerns with the third component of body fixed 
spin operator \cite{bra}. Substantially it is related to the 
orbital angular momentum but no experimental identification has been 
done. 
Therefore in our analysis we put $L=0$ for all multiplets states. 

In YA treatment, one needs to truncate the base in finite size.  
We expand the collective wave function with $N \le 3$, except for 
the states $(S,I)=(-6,0)$ in $\{28\}$, $(-4,0),(-4,1)$ 
and $(-5,1/2)$ in $\{35\}$. In those states, 
we expand the base with $N \le 4$ for obtaining sufficient 
convergence. 

In Fig.\ref{fig3} presents the $\alpha$ dependence of the mass 
spectra within the naive perturbation scheme. 
Actually, we show mass difference between the quantized mass spectra 
and the classical energy. Also Fig.\ref{fig4} presents the results 
of Yabu-Ando treatment. 
In both results the spectra as well as their differences within 
each multiplet decrease monotonically with increasing $\alpha$. 
On the other hand, mass differences between different multiplet but 
with same quantum numbers $(S,I)$ increase, 
which have been already observed in the calculation of $SU(2)$ \cite{myfp}. 
One easily observe that difference between the results of two treatments 
disappears with increasing $\alpha$. This behavior is easily understood. 
In Fig.\ref{fig5} we illustrate the mixing probability of
the multiplet for $NN$ channel in $\{\overline{10}\}$ for various $\alpha$.
For increasing $\alpha$, the mixing of higher representations are
significantly decreased; as a result, the naive perturbation is 
sufficient for the analysis of SU(3) dibaryons for such strong gravity 
region. In our analysis, the pion and the kaon mass and the coupling 
constants are fixed by their experimental values 
(we simply set $F_\kappa=F_\pi$ for the coupling constant) 
and if we take into account variations of the mesonic data about 
change of gravity, exact SU(3) flavor symmetry will attain 
at a strong gravity limit. 

\section{CONCLUSION}
In this Letter, we have studied the gravitational effect 
to the dibaryons in the axially symmetric ES model. 
In particular, we have investigated gravity coupling constant 
dependence of the energy spectra of the SU(3) dibaryons. 
We have used the collective quantization in three flavor space. 
To treat the symmetry breaking term, we employ the lowest order 
(naive) perturbation to that as well as Yabu-Ando treatment.  
Both treatments have shown that mass differences between 
spectra with different strangeness decrease 
monotonically and increase within different multiplet but 
with same quantum numbers $(S,I)$ with increasing $\alpha$.
In the strong gravity limit, the SU(3) flavor 
symmetry recovers; all the spectra degenerate in each multiplet 
and no mixing between the multiplets occur. Such symmetry 
restorations may be observed in high energy experiment at LHC.
 
In this Letter, we treat $\alpha$ as a free parameter. 
In the Einstein-Skyrme theory, the Planck mass is related 
to the pion decay constant $F_{\pi}$ and coupling constant 
$\alpha$ by $M_{pl}=F_{\pi}\sqrt{4\pi /\alpha}$. To realize 
the realistic value of the Planck mass, the coupling 
constant should be extremely small with $\alpha \sim O(10^{-39})$. 
However, we have shown that the effects of gravity can be observed 
only in large $\alpha$. 
Some theories such as scalar-tensor gravity theory~\cite{brans}
and ``brane world scenario''\cite{add} predict large enhancement of the 
gravitational constant (in other words the reduction of 4 dim. Planck mass).  
There may have been an epoch in the early universe and may observe 
at an ultra high energy experiment where the gravitational effects on 
hadrons are crucial. 
 
\begin{acknowledgments}
We would like to thank Rajat K.Bhaduri for drawing our attention to 
this subject and useful comments. Also we deeply appreciate to Noriko Shiiki 
for valuable discussions.  


\end{acknowledgments}



\end{document}